\documentstyle[twocolumn,prl,aps,floats]{revtex}
\input psfig

\def\bea{\begin{eqnarray}}
\def\eea{\end{eqnarray}}
\def\ben{\begin{equation}}
\def\een{\end{equation}}
\def\benu{\begin{enumerate}}
\def\enu{\end{enumerate}}

\def\n{n}

\def\sss{\scriptscriptstyle\rm}

\def\1var{(\bx_1...\bx\N)}

\def\half{\frac{1}{2}}

\def\br{{\bf r}}
\def\bR{{\bf R}}

\def\bx{{x}}

\def\x{_{\sss X}}

\def\s{_{\sss S}}
\def\xc{_{\sss XC}}

\def\N{_{\sss N}}
\def\H{_{\sss H}}

\def\unif{^{\rm unif}}

\def\ee{_{\rm ee}}

\def\sph_int{ {\int d^3 r}}

\input rotate

\begin{document}
\def\siz{}
\siz
\twocolumn[\hsize\textwidth\columnwidth\hsize\csname
@twocolumnfalse\endcsname
\title
{Total energy density as an interpretative tool}
\author{Morrel H. Cohen}
\address
{Department of Physics and Astronomy, Rutgers University,
136 Frelinghausen Road, Piscataway, NJ 08854-8019}
\author{Derek Frydel} 
\address{Department of Chemistry,
Rutgers University,
315 Penn Street,
Camden, NJ  08102}
\author{Kieron Burke}
\address{Department of Chemistry, Rutgers University, 610 Taylor Road, 
Piscataway, NJ 08854}
\author{Eberhard Engel}
\address{Institut f{\"u}r Theoretische Physik, J.W. Goethe-Universit{\"a}t
Frankfurt, Robert-Mayer-Strasse 8-10, D-60054,
Frankfurt am Main, Germany}
\date{In preparation for J. Chem. Phys, \today}
\maketitle
\begin{abstract}
\siz
     We present an unambiguous formulation for the total energy density
within density-functional theory.  We propose that it be used as a tool for
the interpretation of computed energy and electronic structure changes
during structural transformations and chemical reactions, augmenting the
present use of electron density changes and changes in the Kohn-Sham local
density of states and Kohn-Sham energy density.
\end{abstract}
\pacs{71.15.Mb,71.45.Gm,31.15.Ew}
]
\narrowtext

     Changes in the electron density of a system undergoing a structural
transformation or a chemical reaction have long been used to understand the
driving forces underlying the transformation or the reaction.  Indeed, the
electron density itself has been regarded as one of the most significant
descriptors of a system of electrons and nuclei since the advent of quantum
mechanics.  This view of the importance of the electron density was greatly
strengthened by the emergence of density-functional theory.  The
Hohenberg-Kohn theorem\cite{HK64} showed that the total energy of a system could
be treated as a functional of the electron density.   The Kohn-Sham theory\cite{KS65}
makes possible the construction of the true ground-state electron
density of an $N$-electron system from the $N$ one-electron orbitals of lowest
energy of a particular independent-electron system.  This rigorous
decomposition of the electron density into contributions from one-electron
orbitals, the Kohn-Sham orbitals, greatly increases its power as an
interpretive tool.

     At present, methods based on density-functional theory are
dominant in first-principles calculations for condensed matter and for much
of quantum chemistry.  The electron density and its Kohn-Sham orbital
decomposition, fundamental outputs of the computations, are typically used
to enrich understanding of the resulting energetics.  For example, studies
of the isosurfaces of electron density changes along a reaction pathway for
the dissociation of H$_2$ on the Pd (100) surface\cite{WNC99} have revealed the
formation of bridge bonds between the s-p tails of the metallic surface
electron density and the $\sigma_g$ and $\sigma_u^*$  molecular orbitals.
These bridge bonds mediate the hybridization of the molecular orbitals with the metallic
d-orbitals before they actually overlap.  It is the bridge bonds which
evolve into the bonds between the dissociated hydrogen atoms and the metal.

     The Kohn-Sham theory enables one to go still deeper into the
interpretation of the results of the electronic-structure computations.
For extended systems,
one can construct the Kohn-Sham density of states,
\ben
g(\epsilon) = \sum^N_{i=1} \delta (\epsilon-\epsilon_i),
\label{gE}
\een
where $\epsilon_i$ is one of the $N$ lowest Kohn-Sham eigenvalues,
and use it to tease out the energetics associated with the electron-density
changes.  For example, in studies of the interaction of chemisorbed O and H
to form OH and H$_2$O on the (111) surfaces
of Rh and Pt\cite{WNCb99}, the resonances
and bound states associated with the bonding and antibonding orbitals of
the atoms, radicals, or molecules to the surface are shown to be clearly
visible in $g(\epsilon)$.  Even more revealing is the local
density of states,
\ben
g(\br,\epsilon) = 
\sum^N_{i=1} | \phi_i (\br) |^2\, \delta (\epsilon-\epsilon_i),
\label{grE}
\een
where $\phi_i(\br)$ is one of the occupied Kohn-Sham orbitals, which enables one to
associate a particular resonance or bound state in the Kohn-Sham spectrum,
Eq.(\ref{gE}), with a particular atom or molecule.  It allows one to make a
resolution with respect to the Kohn-Sham energy of the changes of electron
density accompanying structural transitions or chemical reactions.

     Nevertheless, what drives a transformation or a reaction is the
dependence of the total energy on the nuclear coordinates.  Thus, what one
needs as an additional interpretive tool is an unambiguous, computationally
feasible total-energy density $e(\br)$, such that
\ben
E = \int d^3r\, e(\br),
\label{etot}
\een
where $E$ is the total energy.  With $e(\br)$ one could explore the relation
between local changes in the electron density and the spatial dispersion of
the total energy changes through the corresponding changes in $e(\br)$.

To be more specific,  in Ref. \cite{WNC99}, evidence was found for the
simultaneous presence of all the specific mechanisms commonly cited as
playing important roles in the dissociation of H$_2$ on a transition
metal surface.  These included the formation of an orthogonality
hole in the $s-p$ tail of the metallic electron density, with concommitant
flow of screening charge into the $d$-states, as emphasized by Harris and
Andersson\cite{HA85}; the hybridization of the $d$-states with the occupied
$\sigma_g$ bonding orbital of the H$_2$, as emphasized by Hammer and
Scheffler\cite{HS95}, and the hybridization of the unoccupied 
$\sigma_u^*$  antibonding  orbital of the H$_2$ with associated
backbonding, as emphasized by Hammer and N{\o}rskov
\cite{HN95,HNb95}.
Each of these processes introduced distinct, easily recognized characteristic
changes in the electron density, which occured in distinct regions of space.
One could therefore utilize $e(\br)$ to order the relative importance
of each of these mechanisms as well as of bridge-bond formation at different
locations along the dissociation pathway, garnering thereby a
detailed, intimate quantitative understanding of the breaking of the
intramolecular bond and the formation of metal-atom bonds in
the course of the dissociation.

     There are several difficulties to be surmounted in constructing an
unambiguous, computationally-feasible expression for $e(\br)$, as will be
discussed below.  To avoid the difficulties, only the Kohn-Sham energy
density $e\s(\br)$,
\ben
e\s(\br)=\sum_{i=1}^N \epsilon_i\, |\phi_i (\br)|^2
= \int^{\epsilon_N^+} d\epsilon\, g(\br,\epsilon)
\label{esr}
\een
was used as an interpretive tool in reference \cite{WNCb99}, 
though to good effect.
In the present paper, we present an explicit formulation for $e(\br)$ which is
both unambiguous and computationally feasible,  as required.

     The total energy can be represented as a density functional in the form 
\ben
E = T + V\ee + V_{en} + V_{nn},
\label{Etot}
\een
where $T$, $V\ee$, and $V_{en}$ are the functionals for the electron kinetic energy,
the electron-electron interaction, and the electron-nuclear interaction.
$V_{nn}$ is the Coulomb interaction between the nuclei. In Kohn-Sham density-functional
theory\cite{DG90,PY89}, the expression Eq. (\ref{Etot}) is rewritten as
\ben
E = T\s + U + E\xc + V_{en} + V_{nn}.
\label{EtotKS}
\een
Here $T\s$ is the kinetic energy of the
independent Kohn-Sham particles,  U
is the Hartree electrostatic energy,
\ben
U = \half \int d^3r\, \int d^3r'\, \frac{n(\br) n(\br')}{|\br-\br'|},
\label{Udef}
\een
where $n(\br)$ is the electron density,
and $E\xc$ is the exchange-correlation
energy,
\ben
E\xc = T - T\s + V\ee - U.
\label{Exc}
\een

On the other hand,
    the total Kohn-Sham energy of the system is
\ben
E\s = \sum_{i=1}^N \epsilon_i = T\s + \int d^3r\, n(\br)\, v\s(\br),
\label{Es}
\een
In Eq.(\ref{Es}),  $v\s(\br)$ is the Kohn-Sham potential,
\ben
v\s(\br) = v_{en}(\br) + v\H (\br) + v\xc(\br),
\label{vs}
\een
where
$v_{en}(\br)$ is the Coulomb potential produced by the nuclei at $\br$,
which enters $V_{en}$ as well
\ben
V_{en} = \int d^3r\, \n(\br)\, v_{en} (\br),
\label{Ven}
\een
$v\H(\br)$ is the Hartree potential, the mean electrostatic potential produced
by the electrons,
\ben
v\H (\br) = \int d^3 r'\, \frac{1}{|\br-\br'|}\, n(\br'),
\label{vH}
\een
which enters the Hartree energy functional,
\ben
U=\int d^3r\, e\H(\br),~~~~e\H(\br)=\half\, n(\br)\, v\H (\br).
\label{eH}
\een
Finally, $v\xc(\br)$ is the functional derivative of $E\xc$:
\ben
v\xc(\br) =\frac{\delta E\xc}{\delta n(\br)}.
\label{vxc}
\een
Thus Eq.(\ref{Es}) can be transformed into
\ben
E\s = T\s + 2U + V_{en}+\int d^3r\, n(\br)\, v\xc(\br).
\label{Esxc}
\een
Inserting Eq.(\ref{Esxc}) into 
Eq.(\ref{Etot}) allows us to eliminate $T\s$, resulting in
\ben
E = E\s - U +V_{nn} + E\xc -\int d^3r\, n(\br)\, v\xc(\br).
\label{Etotxc}
\een
Each of the above terms can unambiguously be written as an integral
over an energy density so that
\bea
e(\br)&=&e\s(\br) - e\H(\br) -
\half \rho_n (\br)\, v_{en}(\br)
\nonumber\\
&+& e\xc (\br) - n(\br)\, v\xc(\br).
\label{er}
\eea
In this expression, $e\s(\br)$ is given by
Eq.(\ref{esr}) and can readily be constructed from the
output of the standard codes which solve the Kohn-Sham equations.
The electron density $\n(\br)$ is a fundamental
output of such computations.  The potentials  $v\H(\br)$
and $v_{en}(\br)$ must be constructed during
the computations, as must be $v\xc(\br)$.
The number density of the
nuclei weighted by their respective
atomic numbers is
\ben
\rho_n (\br) = \sum_\alpha Z_\alpha\, \delta (\br -\bR_\alpha),
\label{rhon}
\een
where $\bR_\alpha$ is the position and $Z_\alpha$ the charge of nucleus $\alpha$.
We note that the
third term in Eq.(\ref{er}), though singular at the nuclei, vanishes between the
nuclei.

The exchange-correlation energy density $e\xc(\br)$ is defined in terms of $v\xc[\n](\br)$ via
the procedure of Burke, Cruz, and Lam (BCL) \cite{BCL98},
an extension of the original ideas of Engel and Vosko\cite{EV93},
\ben
\nabla^2  e\xc(\br)
= 3 \nabla \Big(\n(\br)\ \nabla {\tilde v}\xc (\br)\Big),
\label{excuna}
\een
where
\ben
{\tilde v}\xc[\n](\br) = \int_0^1
\frac{d\gamma}{\gamma}\ v\xc [\n_{\gamma}](\frac{\br}{\gamma}).
\label{vxctildedef}
\een
and
$\n_\gamma (\br)=\gamma^3 \n(\gamma \br)$.
Their definition has several advantages in this context.  Most
importantly, the BCL procedure defines an exchange-correlation
energy density  unambiguously in terms of any given exchange-correlation energy functional
(including the exact one), since it uses the potential.
Thus exchange-correlation energy densities
calculated this way are approximations to an exact quantity.  Within the
local density approximation (LDA), their procedure reproduces simply 
$e\xc\unif (n(\br))$, the exchange-correlation energy density of a uniform
electron gas of density $n(\br)$.  This is just the conventional energy density
within LDA, and so is already calculated in standard LDA Kohn-Sham
calculations.  Within any generalized gradient approximation (GGA) such as
PBE\cite{PBE96} or 
BLYP\cite{B88,LYP88}, their procedure defines a different energy density
from the conventional one.  This energy density includes dependencies on
the Laplacian and other higher derivatives of $n(\br)$, and so is far more
sensitive to details in $n(\br)$ than the conventional GGA forms. 
The scaling defined in Eq. (\ref{vxctildedef}) is easily performed on
any approximate density functional.
Finally, the BCL procedure
defines an exact unambiguous exchange energy density, so that it can be
applied even to hybrids of GGA with exact exchange\cite{BCL98}.

To illustrate this energy density, we plot the various contributions to the
total energy density for the Na atom.  Our calculations are all for exact
exchange only, using the atomic optimized effective potential (OEP) code
of Engel\cite{ED99}.
\begin{figure}
\unitlength1cm
\begin{picture}(12.5,6.5)
\put(-5.2,4.0){\makebox(12,6.5){
\includegraphics{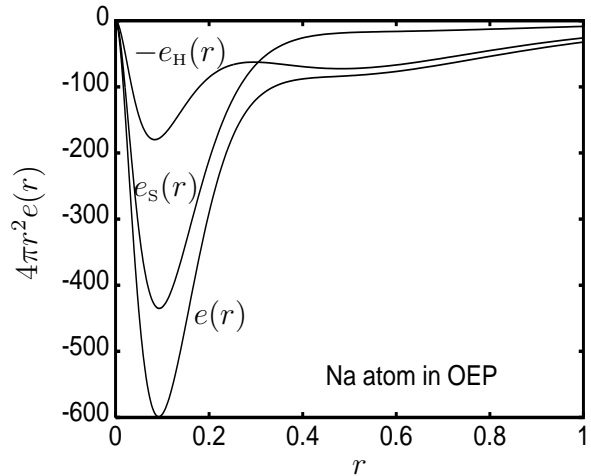}
}}
\setbox6=\hbox{\large $4\pi r^2e(r)$}
\put(0.8,3.85) {\makebox(0,0){\rotl 6}}
\put(5.1,0.5){\large $r$}
\put(3.0,2.5){\large $e(r)$}
\put(2.2,4.2){\large $e\s(r)$}
\put(2.2,6){\large $-e\H(r)$}
\end{picture}
\caption{Radial contributions to the total electronic energy density 
for the Na atom (atomic units).}
\label{Na}
\end{figure}
\begin{figure}
\unitlength1cm
\begin{picture}(12.5,6.5)
\put(-5.2,4.0){\makebox(12,6.5){
\includegraphics{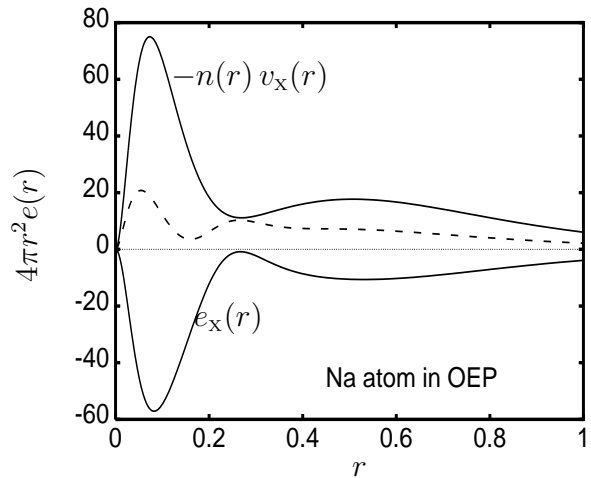}
}}
\setbox6=\hbox{\large $4\pi r^2e(r)$}
\put(0.8,3.85) {\makebox(0,0){\rotl 6}}
\put(5.1,0.5){\large $r$}
\put(3.0,2.5){\large $e\x(r)$}
\put(2.7,5.7){\large $-\n(r)\, v\x(r)$}
\end{picture}
\caption{Radial exchange contributions to the total electronic energy density 
for the Na atom (atomic units); the dashed curve is the sum of the
other two.}
\label{exctcHe}
\end{figure}
From Figs. 1 and 2, we 
see that the total energy density is usually dominated by two terms:
the Kohn-Sham eigenvalue
contribution and the electrostatic energy correction.  The two exchange terms of
Eq. (\ref{er}) are
small, and largely cancel, as shown in Fig. 2.
This cancellation is already apparent in LDA exchange,
where the sum of these terms is only 1/3 of $e\x\unif$.  This agrees with
arguments\cite{Pb78,MA80,HNb97} that changes in the
energy can be largely understood in terms of changes in KS eigenvalues and the electrostatic
potentials.  The present analysis opens the possibility of extending
these arguments to the exact functional.

To illustrate the importance of the electrostatic energy correction, consider
Fig. 3, which shows energy density differences between neutral Na and
its ion.  The Kohn-Sham contribution is dominated by the core 
($r \leq 2$, roughly).  This is because the 3s electron in Na induces an
almost constant shift in the core Hartree potential, so that all eigenvalues are
about 0.29 higher in the neutral relative to the ion.   When the Hartree
correction is made, and the total energy density difference plotted, we
find that there are two almost equal and opposite contributions to the
energy difference.  To see that these are entirely electrostatic effects,
we further subtract $e\H(r)$ from the total (dashed line), showing that,
in the absence of Hartree contributions, the total energy density lies
almost entirely in the valence region.  This simple analysis demonstrates
both the importance of electrostatic contributions to the energy density,
and how the various terms are needed in Eq. (\ref{er}) to produce
the final physical picture.  (For example, overall shifts in KS eigenvalues 
do not contribute to $e(\br)$, but do show up in $e\s(\br)$.)
The electrostatic energy correction is omitted in
the generalized perturbation method\cite{DG76} widely used in the theory
of alloys\cite{M97}.
\begin{figure}
\unitlength1cm
\begin{picture}(12.5,6.5)
\put(-5.2,4.0){\makebox(12,6.5){
\includegraphics{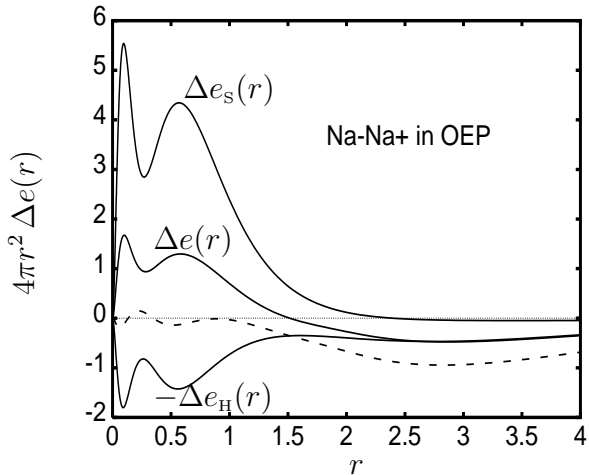}
}}
\setbox6=\hbox{\large $4\pi r^2\, \Delta e(r)$}
\put(0.8,3.85) {\makebox(0,0){\rotl 6}}
\put(5.1,0.5){\large $r$}
\put(2.5,3.5){\large $\Delta e(r)$}
\put(2.9,5.5){\large $\Delta e\s(r)$}
\put(2.5,1.4){\large $ -\Delta e\H(r)$}
\end{picture}
\caption{Same as Fig. 1, but for the difference between the neutral atom, Na,
and the ion, Na$^+$; the dashed curve is $\Delta e(r) - \Delta e\H(r)$.}
\label{Nadiff}
\end{figure}
The present analysis allows us
to explore the role of exhange-correlation energy differences
in chemical systems.

     Note that $v\H(\br)$  and $v_{en}(\br)$ each diverge separately in the
thermodynamic limit for extended systems.  The divergences cancel each
other in $v\s(\br)$, Eq.(\ref{vs}), because only their sum enters.  The divergences
similarly cancel in $E$, Eqs.(\ref{Etot}), (\ref{EtotKS}), and 
(\ref{Etotxc}) and in $E\s$, Eq.(\ref{Es}).  They
do not cancel in $e(\br)$ as written in Eq.(\ref{er}).
This difficulty
should be obviated by the methods of eliminating the singular parts of
$v\H(\br)$  and $v_{en}(\br)$  which are standard in the theory of the electron
structure of extended systems.  That is, $v\H(\br)$  and $v_{en}(\br)$ should be
interpreted as the nonsingular part of the Hartree and nuclear potential,
respectively, in Eq.(\ref{er}) for extended systems.

     In conclusion, we propose (1.) that the total energy density between
two nuclei
be constructed from the output of Kohn-Sham computations augmented by the
procedure of ref. \cite{BCL98} and used as an interpretive tool in analyzing energy
and electronic structure changes during transformations and reactions when
the focus is on the study of bond formation in the spaces between atoms.
We propose in addition (2.) that the full expression for $e(\br)$, Eq.(\ref{er}), be
coarse-grained by integration over a small sphere containing a particular
nucleus to examine the role of the corresponding atom in chemical bonding
or structural transformation.  Such spheres are already present in, e.g.,
linear-augmented-plane-wave, muffin-tin, and KKR codes and can be readily
introduced into other procedures.

\section{Acknowledgment}
     This work was initiated at the Aspen center of Physics and carried
forward at the Department of Physics of Rutgers University (New Brunswick).
D.F. and K.B. were supported by a grant from Research Corporation, and
the
National Science Foundation under grant number 
We thank Eberhard Engel for the use of his atomic OEP code.


\begin{references}
\siz
\bibitem{HK64}
{\em Inhomogeneous electron gas},
P. Hohenberg and W. Kohn, Phys. Rev. {\bf 136}, B 864 (1964).

 
\bibitem{KS65}
{\em Self-consistent equations including exchange and correlation effects},
W. Kohn and L.J. Sham, Phys. Rev.  {\bf 140}, A 1133 (1965).

 
\bibitem{WNC99}
{\em Mechanisms in the dissociation of H$_2$ on Pd (100)},
S. Wilke, V. Natoli, M. H. Cohen, in preparation.

 
\bibitem{WNCb99}
{\em Theoretical investigation of water formation on Rh and Pt surfaces},
S. Wilke, V. Natoli, M. H. Cohen, in preparation.

 
\bibitem{HA85}
{\em H$_2$ dissociation at metal surfaces},
J. Harris and S. Andersson, Phys. Rev. Lett. {\bf 55}, 1583 (1985).

 
\bibitem{HS95}
{\em Local Chemical Reactivity of a Metal Alloy Surface},
B. Hammer and M. Scheffler, Phys. Rev. Lett. {\bf 74}, 3487 (1995).

 
\bibitem{HN95}
{\em Why gold is the noblest of all the metals},
B. Hammer and J.K. N{\o}rskov, Nature {\bf 376}, 238 (1995).

 
\bibitem{HNb95}
B. Hammer and J.K. N{\o}rskov, Surf. Sci. {\bf 343}, 211 (1995).


 
\bibitem{DG90}
R.M.  Dreizler  and  E.K.U.  Gross,    {\sl Density   Functional   Theory }
    (Springer-Verlag, Berlin, 1990).

 
\bibitem{PY89}
{\em Density  Functional  Theory  of  Atoms  and Molecules}, R.G. Parr and W. Yang 
(Oxford, New York, 1989).

 
\bibitem{BCL98}
{\em Unambiguous exchange-correlation energy density},
K. Burke, \underline{F.G. Cruz}, and \underline{K.C. Lam}, J. Chem. Phys. {\bf 109}, 8161 (1998).

 
\bibitem{EV93}
{\em Exact exchange-only potentials and the virial relation as microscopic criteria for generalized gradient approximations},
E. Engel and S.H. Vosko, Phys. Rev. B {\bf 47}, 13164 (1993).

 
\bibitem{PBE96}
{\em Generalized gradient approximation made simple},
J.P.~Perdew, K.~Burke, and M.~Ernzerhof, Phys. Rev. Lett. {\bf 77}, 3865 
(1996); {\bf 78}, 1396 (1997) (E).
 
\bibitem{B88}
{\em Density-functional exchange-energy approximation with correct asymptotic behavior},
A.D. Becke, Phys. Rev. A {\bf 38}, 3098 (1988).

 
\bibitem{LYP88}
{\em Development of the Colle-Salvetti correlation-energy formula into a functional of the electron density},
C. Lee, W. Yang, and R.G. Parr, Phys. Rev. B {\bf 37}, 785 (1988).

 
\bibitem{ED99}
{\em From explicit to implicit density functionals}, 
E. Engel and R.M. Dreizler, J. Comput. Chem. {\bf 20}, 31 (1999).

 
\bibitem{Pb78}
{\em Individual orbital contributions to the SCF virial in homonuclear diatomic molecules},
D.G. Pettifor, J. Chem. Phys. {\bf 69}, 2930 (1978).

 
\bibitem{MA80}
{\em The electronic structure of transition metals}, 
A.R. Mackintosh and O.K. Andersen, in {\em Electrons at the Fermi Surface}, ed. M. Springford, (Cambridge University Press,
Cambridge, 1980).
 
\bibitem{HNb97}
{\em Theory of adsorption and surface reactions}, B. Hammer and J.K. N\"orskov, in 
{\em Chemisorption and reactivity on supported clusters and thin films}, eds. R.M. Lambert and G. Pacchioni (Kluwer, Holland,
1997).
 
\bibitem{DG76}
{\em  Generalized perturbation theory in disordered transitional alloys--applications to calculation of ordering
energies},
F. Ducastelle and F. Gautier, J. Phys. F {\bf 6}, 2039 (1976).
 
\bibitem{M97}
{\em First principles approaches to surface segregation},
R. Monnier, Phil. Mag. B, {\bf 75}, 67 (1997).

 
\end{references}
\end{document}